\documentclass[10pt,letterpaper,twocolumn]{article}

\usepackage[letterpaper,       
            top=1.91cm,        
            bottom=2.54cm,     
            left=1.91cm,       
            right=1.91cm,      
            columnsep=0.84cm,  
            headheight=37pt,headsep=12pt]{geometry}
\usepackage[T1]{fontenc}
\usepackage[utf8]{inputenc}
\usepackage{lmodern}
\usepackage{microtype}
\usepackage{etoolbox}

\usepackage[sb]{libertine} 
\usepackage[varqu,varl]{inconsolata} 
\usepackage{iftex}
\ifPDFTeX\else
\fi

\usepackage{xcolor}
\definecolor{ttorange}{HTML}{FF5000}      
\definecolor{ttorangeD}{HTML}{FF5000}     

\definecolor{ttaccent}{HTML}{FFD54F}
\definecolor{ttink}{HTML}{1A1A1A}
\definecolor{ttgray}{HTML}{5A5A5A}
\definecolor{ttbg}{HTML}{FFF6EC}
\definecolor{ttline}{HTML}{EAD9C6}

\usepackage{graphicx}
\graphicspath{{./}{./assets/}}

\usepackage{amsmath,amssymb,amsthm}
\usepackage[libertine]{newtxmath}
\usepackage{booktabs}
\usepackage{multirow}
\usepackage{caption}
\theoremstyle{plain}

\theoremstyle{definition}

\theoremstyle{remark}

\usepackage{titlesec}

\usepackage{enumitem}
\setlist[itemize]{leftmargin=1.2em,itemsep=1pt,topsep=2pt}
\setlist[enumerate]{leftmargin=1.4em,itemsep=1pt,topsep=2pt}

\usepackage{titlesec}
\titleformat{\section}
  {\normalfont\large\bfseries\color{ttink}}
  {\textcolor{ttorange}{\thesection}}{0.55em}{}
\titleformat{\subsection}
  {\normalfont\normalsize\bfseries\color{ttink}}
  {\textcolor{ttorange}{\thesubsection}}{0.55em}{}
\titleformat{\subsubsection}
  {\normalfont\normalsize\itshape\color{ttink}}
  {\thesubsubsection}{0.5em}{}
\titlespacing*{\section}{0pt}{9pt plus 2pt minus 1pt}{6pt}
\titlespacing*{\subsection}{0pt}{7pt plus 1pt minus 1pt}{3pt}

\usepackage{fancyhdr}
\fancypagestyle{reportfirst}{%
  \fancyhf{}
  \fancyhead[L]{\includegraphics[height=25pt]{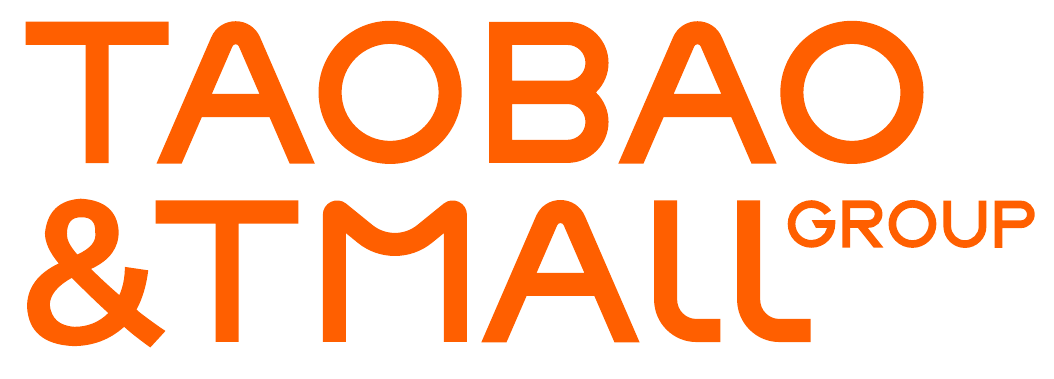}}
  \fancyfoot[C]{\scriptsize\color{ttgray}\thepage}
  
  \renewcommand{\headrule}{\hbox to\headwidth{\color{ttline}\leaders\hrule height 0.4pt\hfill}}
}

\usepackage[numbers,sort&compress]{natbib}
\usepackage{adjustbox}
\usepackage{balance}
\usepackage{url}
\usepackage[colorlinks=true,linkcolor=ttorange,citecolor=ttorange,
            urlcolor=ttorange]{hyperref}

\usepackage{tcolorbox}
\tcbuselibrary{breakable}

\newcommand{\reporttitle}[1]{\def\thereporttitle{#1}}
\newcommand{\reportauthors}[1]{\def\thereportauthors{#1}}
\newcommand{\authormark}[2]{{\small #1}\textsuperscript{\color{ttgray}#2}}

\newtcolorbox{promptbox}[1][]{%
  colback=gray!5!white,colframe=gray!75!black,fonttitle=\bfseries,
  title=#1,boxrule=0.5pt,arc=2pt,left=4pt,right=4pt,top=4pt,bottom=4pt,
  fontupper=\small\ttfamily,breakable=false
}

\begin{document}
\raggedbottom
\reporttitle{From Visual Similarity to Exact Match: Joint Mining of Product and User Interactions for VLM Ranking in E-commerce Multimodal Search}
\reportauthors{%
  \mbox{\authormark{Qing Liu}{*}, \authormark{Lianyu Du}{*},
  \authormark{Qiuyu Zhao}{*}, \authormark{Zenghui Sun}{},
  \authormark{Xu Chen}{}, \authormark{Jinsong Lan}{},
  \authormark{Xiaoyong Zhu}{}, \authormark{Bo Zheng}{\dag}}\par
  \vspace{6pt}
  {\small\textbf{Taobao \& Tmall Group of Alibaba}\par}
  \vspace{4pt}
  {\small\color{ttgray}\textsuperscript{*}Equal contribution.\quad
   \textsuperscript{\dag}Corresponding author.}
}

\twocolumn[{%
  \begin{center}
    \vspace*{10pt}
    {\huge\bfseries\color{ttink}\thereporttitle\par}
    \vspace{16pt}
    \thereportauthors
    \vspace{8pt}
  \end{center}
  \begin{center}
  \begin{tcolorbox}[width=0.92\textwidth,colframe=ttorange,
    colback=ttorange!3!white,arc=2mm,boxrule=0.8pt,
    left=35pt,right=35pt,top=22pt,bottom=22pt,boxsep=0pt]
    {\centering\large\bfseries\color{ttorange}Abstract\par}
    \vspace{8pt}
Image-based product search often retrieves visually similar items that differ in identity-critical specifications. Existing methods optimize visual similarity or reason over predefined attributes, but rarely learn item-specific, behavior-informed representations for query interpretation and ranking. We represent decision-relevant attributes as schema-valid keys and catalog-supported values selected using behavioral signals. Query-product agreement on these attributes provides factor-level compatibility evidence for exact-match reranking. We propose a three-stage framework. First, an offline teacher uses product content and privileged evidence from historical search and filtering interactions to train a content-only student, which infers behavior-informed attributes without interaction features at serving time. These attributes are validated against multiple catalog sources for compatibility scoring. Second, a retrieval-augmented multimodal large language model performs schema-grounded visual reasoning to predict visually supported query attributes in the same catalog schema. Third, an offline ranking policy adapts attribute weights and hierarchical fusion priorities to query intent, aligning queries and products through visual fidelity and factor-level compatibility. On a large-scale e-commerce platform, our framework improves human-evaluated Top-1 exact-match relevance by 5.54 percentage points. Online A/B testing increases transactions by 2.16\% and GMV by 4.11\%, demonstrating the effectiveness of behavior-informed product representations, query-side visual reasoning, and SKU-level ranking.
    \par\vspace{12pt}
    {\footnotesize\color{ttgray}
      \par
      \vspace{3pt}
      \mbox{\textbf{Correspondence:}
      {\color{ttorange}\{\href{mailto:liuqing.liu@taobao.com}{liuqing.liu},
      \href{mailto:dulianyu.dly@taobao.com}{dulianyu.dly},
      \href{mailto:zhaoqiuyu.zqy@taobao.com}{zhaoqiuyu.zqy}\}@taobao.com}\textsuperscript{*};
      \href{mailto:bozheng@alibaba-inc.com}{bozheng@alibaba-inc.com}
      \textsuperscript{\dag}}\par}
  \end{tcolorbox}
  \end{center}
  \vspace{10pt}
}]
\thispagestyle{reportfirst}
\section{Introduction}
In visual search scenarios, where consumers upload images to find matching products, existing approaches primarily rely on deep visual representations to retrieve items with similar appearances \cite{clip, jia2021, siglip,revision}. While effective for general retrieval tasks \cite{chen2020simple}, these methods face significant challenges in modern e-commerce. Specifically, online marketplaces are increasingly saturated with visually homogeneous products—items that share strikingly similar aesthetics yet differ fundamentally in critical attributes such as material quality, functional design, or brand authenticity. 

\begin{figure}[t]
  \centering
  \includegraphics[width=0.95\linewidth]{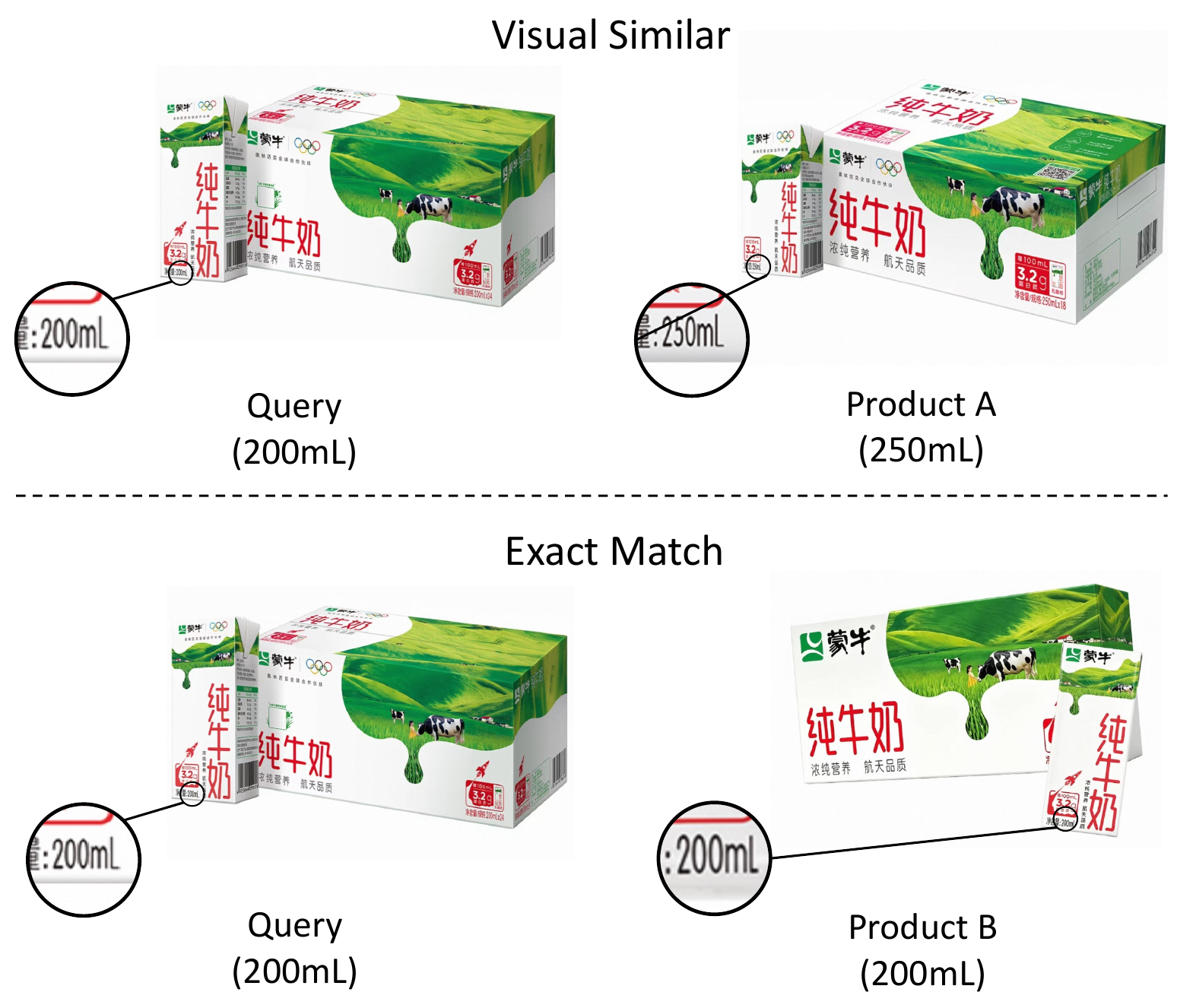}
    \caption{
   Example illustrating visual similarity versus exact match. The query product (200 mL) and Product A (250 mL) have similar packaging but differ in volume, whereas Product B matches the query volume of 200 mL.}
\label{fig:case1}
\end{figure}

As illustrated in Figure~\ref{fig:case1}, the query product and Product A look almost identical, yet they correspond to different variants because the package volume is different (200 mL vs. 250 mL).
In contrast, exact matching requires agreement on identity-critical specifications despite viewpoint or layout differences, as both the query and Product B show 200 mL.
Consequently, systems driven chiefly by global visual similarity often fail to discern these fine-grained distinctions, returning results that are visually plausible but practically misaligned with user needs. 

As a result, users often abandon the initial set of recommendations, resorting to manual exploration across multiple pages, applying filters, and cross-checking details before making a decision. This disconnect between system output and actual intent highlights a fundamental limitation in current approaches: the lack of deep semantic reasoning beyond appearance. Exact matching in e-commerce faces three critical challenges:

\begin{enumerate}
    \item \textbf{Attribute Utility under Sparse Evidence}: Multi-source catalog attributes may be redundant or conflicting, while item-level search and filtering evidence is sparse and exposure-biased. The product representation must therefore separate which keys are decision-relevant from which candidate values are factually valid.
    \item \textbf{Query-Side Schema Grounding}: A query image may not visibly expose every specification, and unconstrained generation can produce plausible but catalog-unsupported attributes. Query interpretation must be grounded in a product-relevant schema and candidate-value space.
    \item \textbf{Intent-Dependent Ranking Priorities}: Visual similarity, factor compatibility, and personalized relevance can disagree, and the relative importance and fusion order of factors vary across query intents. A fixed weighting or priority order is therefore insufficient.
\end{enumerate}
To address this issue, we introduce \textbf{schema-grounded visual reasoning REM (Reasoning-Enhanced Exact Match)}, a simple framework designed to inject fine-grained semantic interpretation into the major stages of visual search. Our framework consists of the following components:
\paragraph{\textbf{Offline Behavior-Informed Decision-Relevant Attribute Learning:}} We formulate decision-relevant attributes as an executable selection over the product's multi-source key-value inventory. A teacher LLM reconciles merchant-curated product identity signals with two-level item/category search and filtering evidence, jointly estimating attribute-key utility and candidate-value validity. Its structured outputs supervise a content-only student through generative SFT, enabling catalog-scale inference on head, tail, and zero-behavior products without behavioral input at serving time.
\paragraph{\textbf{Retrieval-Augmented Schema-Constrained MLLM Prediction:}} We equip the MLLM with a retrieval-guided constraint mechanism. By fetching product-grounded schemas and decision-relevant candidate sets as context, we constrain the model's output space to valid attribute values. This candidate-aware prediction supports schema-consistent attribute recognition by selecting from retrieved candidates rather than generating plausible but potentially incorrect descriptions.
\paragraph{\textbf{Hybrid Heterogeneous Reranking:}} Online reranking is cast as Query-Doc factor compatibility alignment. An offline ranking policy---generated per query-intent cluster from mined decision-relevant factors---modulates AVF-IPF factor weights and selects the priority order of a hierarchical tiered fusion, bridging visual fidelity and factor-level compatibility without additional online model calls.

We have deployed \textbf{REM} in the Taobao mobile application's visual search system \cite{pailitao}. We evaluate product-side supervision through controlled teacher-target variants and behavior-stratified analysis, query-side grounding through schema-constrained attribute prediction, and reranking through offline policy ablation, human relevance judgments, and two-week online A/B tests. Together, these experiments test the three-stage evidence chain from behavior-informed product representation to factor-aware ranking.
The three modules are decoupled from the initial retrieval system, allowing REM to be integrated into an existing search pipeline without replacing its retrieval components. This modular design also supports deployment across heterogeneous e-commerce categories.


\section{Related Work}
\begin{figure*}[t]
  \centering
  \vspace{-9pt}
  \includegraphics[width=0.99\linewidth]{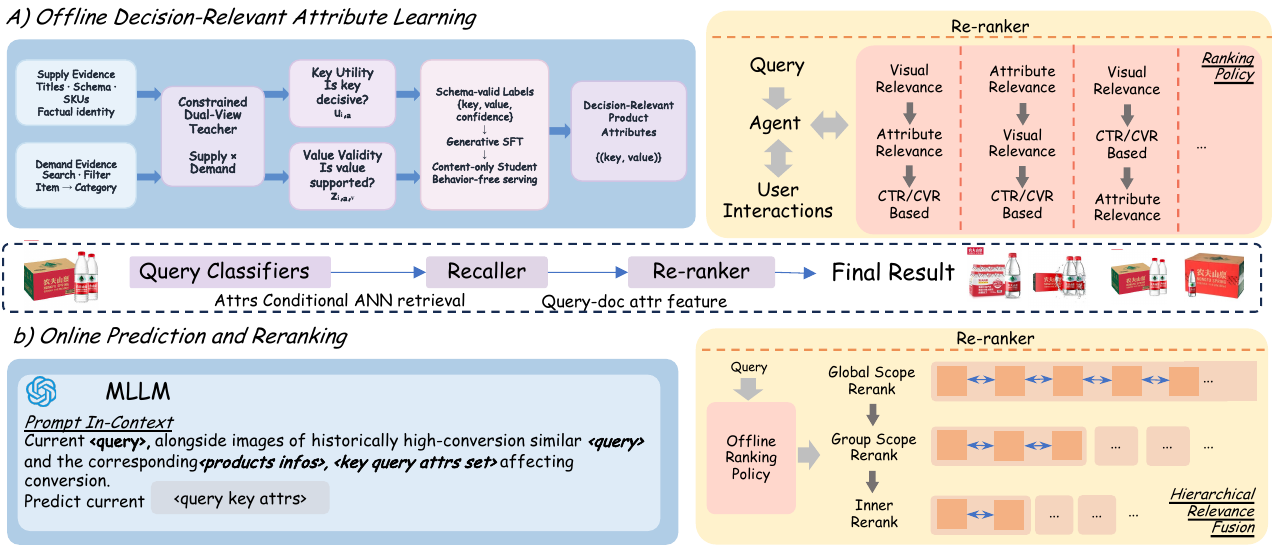}
  \vspace{-0.3cm}
    \caption{Overview of REM. Offline behavior-informed decision-relevant attribute learning, online retrieval and prediction, and reranking operate alongside the existing search components shown in purple.}

\label{pipeline_1}
\end{figure*}
\paragraph{\textbf{Product Attribute Understanding and Decision-Relevant Representation}}
Early product attribute prediction methods primarily relied on multi-task convolutional networks to learn shared representations across predefined attributes \cite{hand2016attributes,lu2017fully}. Subsequent studies introduced part-based models and weakly supervised attention to recognize fine-grained visual properties \cite{tang2019improving,li2015multi}. More recently, vision-language models have enabled open-vocabulary attribute recognition through cross-modal representation learning \cite{ovarnet,ovad,bind}, while large language models have also been studied for data-centric feature selection \cite{li2024exploring}. These methods primarily address attribute recognition or generic feature selection rather than behavior-informed, schema-valid product representations.

Our problem is different. Exact-match search does not require all recognizable attributes to contribute equally; it benefits from a selected subset that captures product and purchase-decision distinctions while remaining compatible with downstream matching. We therefore study behavior-informed decision-relevant attributes over an existing product schema. Merchant titles and SKU structures provide supply-side salience and identity evidence, whereas search and filtering logs provide demand-side evidence of the dimensions users actively express or refine. Using behavior during supervision but not serving follows the learning-using-privileged-information setting, where richer training-time behavioral evidence is unavailable during downstream inference \cite{vapnik2009new}; transferring the teacher's structured decisions to a content-only student is related to generalized distillation \cite{lopezpaz2016unifying}.  REM does not introduce a new distillation paradigm: its contribution is to reconcile the two evidence views into schema-valid key-utility and value-validity targets and evaluate their transfer across behavior strata.
\paragraph{\textbf{Multimodal Representation}}
Early works focused on multimodal embeddings that fuse visual features with textual information from product titles or descriptions \cite{siddiquie2011image}. \citet{collell2017imagined} utilize a language-to-vision mapping and use the predicted visual features to build multimodal representations. Models such as CLIP and SigLIP \cite{clip,siglip} primarily employ contrastive learning to align image and text embeddings. Attribute-aware representations have been studied for fashion retrieval \cite{rubio2017multi}, while multi-intent attribute matching has been studied in e-commerce search \cite{li2024multi}. Kenan E Ak et al. \cite{ak2018learning} jointly predict product categories and attributes while learning discriminative embeddings; VLM2Vec~\cite{vlm2vec,vlm2vec2} uses task-specific instructions to learn general-purpose multimodal embeddings. Retrieval-augmented generation grounds model outputs in retrieved external evidence \cite{rag}. REM retrieves a product-relevant catalog schema and candidate values, then uses the MLLM to select factors in the same space used for reranking. This grounds query interpretation in the schema rather than free-form generation or representation alignment.

\paragraph{\textbf{Multimodal Reranking}}

Reranking is a critical post-retrieval stage that refines an initial candidate set into a final ranked list aligned with user intent. Industrial product search has studied scalable representation-based retrieval \cite{li2021taobao}, while \citet{li2021albef} demonstrate that aligning visual and textual representations before fusion improves cross-modal understanding.
\citet{cui2022m6rec} extend multimodal pretraining to open-ended recommendation, leveraging generative language models augmented with visual signals to improve ranking across diverse product categories. \citet{2025multimodal} identify modality misalignment between independently trained text and visual encoders as a key bottleneck in product search under scalable dot-product scoring constraints, proposing joint alignment objectives to bridge this gap.

Prior work has largely focused on fixed catalog factors, leaving their intent-specific selection, weighting, and ordering less explored. REM identifies item-specific, schema-valid factors, scores matches by rarity and discriminability, and learns intent-specific fusion policies offline for deterministic online execution.
\section{Methods}
Figure~\ref{pipeline_1} illustrates the complete three-stage REM pipeline. The first stage learns a product-side behavior-informed decision-relevant attribute representation for subsequent query-product alignment. A teacher LLM reasons over a fixed multi-source attribute inventory using two complementary views: supply-side salience in product and SKU titles, and demand-side salience from search and filtering behavior. It assesses the decision utility of candidate keys and the validity of their values, producing structured, schema-grounded supervision for a content-only student. The second stage then retrieves product-grounded decision-relevant attribute schemas and guides an MLLM to predict schema-consistent query attributes from query images. The third stage finally casts reranking as Query-Doc factor-compatibility alignment: treating the learned attributes as decision-relevant ranking factors, an offline ranking policy adapts their weights and hierarchical fusion priority per query intent, jointly bridging visual similarity and factor-level compatibility.

\subsection{Behavior-Informed Decision-Relevant Attributes}
Let \(\mathcal{A}_{i}^{\mathrm{ms}}=\{(a,v,s)\}\) denote the multi-source attribute inventory of product \(i\), where \(a\), \(v\), and \(s\) represent an attribute key, a candidate value, and its source, respectively. Its executable key space is \(\mathcal{K}_{i}=\{a\mid(a,v,s)\in\mathcal{A}_{i}^{\mathrm{ms}}\}\). We define the behavior-informed decision-relevant attribute representation as a selected subset of the inventory that captures product and purchase-decision distinctions useful for factor-level compatibility matching. It separates two complementary decisions: the binary key-utility decision \(u_{i,a}\in\{0,1\}\) indicates whether attribute dimension \(a\) is decision-relevant, while \(z_{i,a,v}\in\{0,1\}\) indicates whether candidate value \(v\) is factually supported by the product. The executable representation is
\begin{equation}
\mathcal{A}_{i}^{\mathrm{dec}}
=\{(a,v)\mid(a,v,s)\in\mathcal{A}_{i}^{\mathrm{ms}},
u_{i,a}=1,z_{i,a,v}=1\}.
\label{eq:decision_relevant_constraint}
\end{equation}
This representation suppresses irrelevant or conflicting catalog information while retaining a selected set of key-value pairs that can be directly consumed as factor-level compatibility signals in downstream matching. The multi-source inventory defines the candidate space, but does not itself establish either decision importance or factual validity. We learn this representation through two-stage \textbf{Progressive Attribute Reasoning (PAR)}.

\paragraph{\textbf{Stage 1: Supply--Demand Evidence Construction.}}
Decision relevance is typically latent rather than directly observable. We therefore construct two complementary views of attribute salience. \textbf{Supply-side evidence} is derived from product and SKU titles. Besides expressing product identity and variant facts, these merchant-curated descriptions reflect which properties merchants choose to expose within limited display space. Product titles provide item-level salience, while SKU titles reveal dimensions that distinguish purchasable variants. As titles may also contain promotional, optimization, or templated content, we treat them as evidence rather than ground-truth labels.

\textbf{Demand-side evidence} is estimated at the attribute-key level, with search and filtering retained as separate channels. Text queries entering the platform-wide product search engine are parsed by the production named-entity recognition system, and the extracted behavioral keys are associated with corresponding exposed products. When a user selects a filtering condition \(b=v\), products uniquely exposed immediately before that action receive one filtering exposure and one observation for behavioral key \(b\). Behavioral observations are assigned only to keys as they reveal interest in an attribute dimension yet lack reliable product-value supervision.

For channel \(x\in\{s,f\}\), let \(C_{\ell,b}^{x}\) denote the observations of behavioral key \(b\) and \(E_{\ell}^{x}\) the eligible exposures at item or category level \(\ell\). We first estimate the category-level rate and then smooth sparse item-level evidence toward the corresponding category:
\begin{equation}
\widehat{p}_{c,b}^{x}
=\frac{C_{c,b}^{x}}{E_{c}^{x}+\epsilon},\quad
\widetilde{p}_{i,b}^{x}
=\frac{C_{i,b}^{x}+\kappa_{i}^{x}\widehat{p}_{c,b}^{x}}
{E_{i}^{x}+\kappa_{i}^{x}},
\label{eq:hierarchical_behavior}
\end{equation}
where we set \(\epsilon=1\) to prevent division by zero, and \(\kappa_{i}^{x}\) controls the strength of the category prior. Reliable item evidence dominates for active products, whereas sparse or zero-behavior items back off only to evidence from their product category. We retain at most the Top-30 keys at the item and category levels for each channel, together with their scores, observation counts, exposure counts, and source levels.

\paragraph{\textbf{Stage 2: Constrained Dual-View Reasoning and Distillation.}}
The teacher LLM receives the product title, SKU titles, category path, multi-source key-value inventory with source provenance, and two-level item/category search and filtering evidence. Before utility estimation, it semantically grounds each behavioral key \(b\) to a candidate schema key \(a\in\mathcal{K}_{i}\) or to \textsc{NIL}. This handles lexical variants such as ``memory'' and ``memory capacity,'' while using product and category context to resolve ambiguous mappings such as runtime memory versus device storage.

The grounded evidence is reconciled through two coupled tasks. First, the teacher makes a binary key-utility decision \(u_{i,a}^{T}\in\{0,1\}\) from merchant-side salience, product-identity or SKU-variant relevance, and user-side behavioral support. Second, it predicts value validity \(z_{i,a,v}^{T}\) from the product title, SKU structure, and cross-source consistency. This separation allows behavior to affect which dimensions matter without incorrectly treating it as value supervision. It also preserves multiple values when they represent legitimate product variants while removing values that conflict with all available product evidence.

For each \(a\in\mathcal{K}_{i}\), the teacher outputs a binary \texttt{selected} decision, represented by \(u_{i,a}^{T}\). For every associated candidate value, it also predicts a validity label \(z_{i,a,v}^{T}\) with a structured evidence summary. Key selection is defined as
\begin{equation}
\mathcal{D}_{i}^{T}=\{a\in\mathcal{K}_{i}\mid u_{i,a}^{T}=1\}.
\label{eq:teacher_selection}
\end{equation}
The final teacher output follows Eq.~\ref{eq:decision_relevant_constraint}, using \(\mathcal{D}_{i}^{T}\) and \(z_{i,a,v}^{T}\). The key constraint ensures that every output is matchable downstream, while value validation removes factual conflicts without collapsing legitimate multi-SKU values. Potentially important keys expressed in titles or behavior but absent from the inventory are logged as schema gaps rather than selected outputs.

Behavioral evidence is privileged information because it is sparse, delayed, or unavailable for much of the catalog. We serialize the teacher's key-utility assessments, value-validity decisions, and final decision-relevant key-value set as structured textual targets. A student model is then fine-tuned autoregressively on these targets while observing only \(X_i^S=\{T_i,S_i,C_i,\mathcal{A}_{i}^{\mathrm{ms}}\}\), where \(T_i\), \(S_i\), and \(C_i\) denote the product title, SKU titles, and category path. At inference time, the student generates the same structured decision representation without behavioral input. This transfers supply--demand decision regularities into content-based predictions and enables catalog-scale inference for head, tail, and zero-behavior products.
\subsection{Retrieval-Augmented MLLM Prediction}
Leveraging the decision-relevant attributes already mined on the product side, we employ an MLLM to predict corresponding query-image attributes at inference time and use RAG to improve schema consistency between the query and candidate products.

\paragraph{\textbf{Retriever Design.}} We utilize historical user transaction signals as supervision. For each historical transaction query \( q_{\text{hist}} \) in search logs, we aggregate the decision-relevant attributes from top-K retrieved items to construct an attribute candidate set \( \mathcal{C}(q_{\text{hist}}) \). This process forms a knowledge base \( \mathcal{K} \) that maps embeddings produced by the production query-representation model, \( E(q_{\text{hist}}) \), to their corresponding attribute sets. Unlike non-RAG models that overemphasize visually salient features (e.g., color, sleeve length), the retrieved candidates in \(\mathcal{K}\)
provide schema-grounded priors for weakly visual factors such as brand, model, and specifications.

\paragraph{\textbf{Schema-Constrained MLLM Prediction.}} Generic attribute prediction methods can be misaligned with downstream tasks and generate unsupported values. Our approach instead retrieves product-specific schema information and constrains the MLLM to select values from schema-defined candidate sets, restricting predictions to catalog-supported attributes. For a new image query \( q_{\text{new}} \), the system retrieves historical context by semantic similarity:
\begin{equation}
\label{eq:query_hist}
\mathcal{N}_{10}(q_{\text{new}})
=\operatorname{Top}_{10,\,q_{\text{hist}}\in\mathcal{K}}
\operatorname{sim}\!\left(E(q_{\text{new}}),E(q_{\text{hist}})\right).
\end{equation}
The candidate sets associated with these 10 historical queries are aggregated in \(\mathcal{A}_{\text{dyn}}=\operatorname{Aggregate}\{\mathcal{C}(q):q\in\mathcal{N}_{10}(q_{\text{new}})\}\) and injected into the in-context information. This transforms the MLLM's inference task from open-ended attribute generation to constrained selection within a dynamic, small-scale candidate set:
\begin{equation}
\label{eq:final_attr}
A_{\text{final}} = \text{MLLM}(q_{\text{new}}, \text{context}=\mathcal{A}_{\text{dyn}})
\end{equation}
This construction aligns query-side prediction with the document-side schema and improves the consistency of attribute recognition. 

\paragraph{\textbf{Model training.}} We perform supervised fine-tuning on the Qwen3-VL-2B backbone \cite{qwen3vl} to adapt it for schema-constrained attribute selection. To mitigate hallucination and improve robustness, we introduce candidate dropout during training. Specifically, we randomly mask a subset of candidate values in the input context to simulate incomplete or noisy retrieval. This encourages the model to use cross-modal evidence rather than superficial keyword matching when selecting among the remaining candidates. We use standard cross-entropy loss.
\begin{equation}
    \mathcal{L}_{\mathrm{SFT}} = -\frac{1}{N} \sum_{i=1}^{N} \sum_{t=1}^{T_i} \log P(x_t^{(i)} \mid x_{<t}^{(i)}),
\end{equation}
 where $x_{<t}^{(i)}$ denotes the context before position $t$ in the $i$-th sequence, and $P(x_t^{(i)} \mid x_{<t}^{(i)})$ is the predicted probability of the true token position.

\subsection{\textbf{Hybrid Heterogeneous Reranking}}
\begin{figure}[t]
  \centering
  \includegraphics[width=\linewidth]{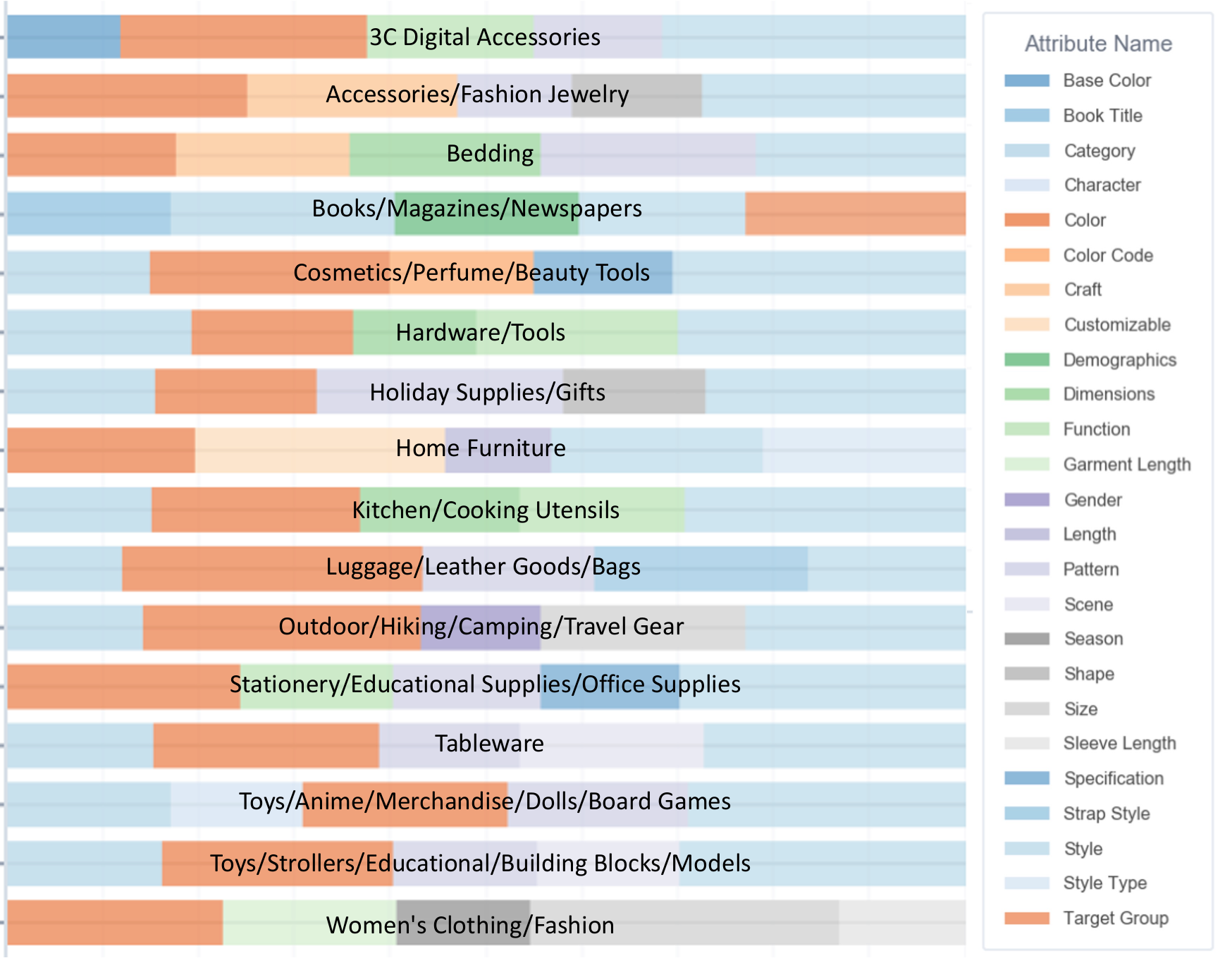}
  \caption{Distribution of decision-relevant attributes across sampled categories. Bar area denotes influence weight, illustrating that purchasing factors vary across categories.}
\label{case_1}
\end{figure}

As visually similar products proliferate in e-commerce catalogs, users must browse multiple detail pages and manually apply filters to locate the exact item they seek, increasing cognitive cost and reducing conversion.
Bridging the gap from \textit{visual similarity} to \textit{exact match} requires reasoning over decision-relevant factors (e.g., material, specification, volume, and brand) whose relative importance is category and intent-dependent (Figure~\ref{case_1}).

We formulate online reranking as \textbf{Query-Doc factor compatibility alignment}, which leverages the decision-relevant factors mined offline to promote items agreeing on intent-relevant catalog dimensions. Because factor importance and fusion priority vary across categories, no single static configuration suffices. We therefore introduce an \textbf{Offline Ranking Policy}, generated once per query-intent cluster during offline mining and executed online through an expected-constant-time hash-map lookup followed by deterministic sorting. The policy \emph{modulates} the scoring and fusion components by adjusting factor weights and tier priorities, while retaining the original static configuration as a fallback.

\paragraph{\textbf{Decision-Relevant Factor Weighting.}}
\label{sec:attribute_weighting}
The decision-relevant factors mined offline exhibit heterogeneous distributions across categories (Figure~\ref{case_1}): matching a rare factor (e.g., \texttt{material:cashmere}) provides more discriminative evidence than matching a ubiquitous one (e.g.,\texttt{sleeve\_style:short\_sleeve}). We quantify this distinction through \textbf{AVF-IPF} (Attribute-Value Frequency--Inverse Product Frequency), which scores each factor-value pair $(a,v)$ by the product of global rarity and query-local discriminability.

\paragraph{\textbf{Inverse Product Frequency (IPF):}} IPF captures \textit{catalog-level rarity}:
\begin{equation}
\text{IPF}(a, v) = \log\frac{|\mathcal{P}| + 1}{N((a, v), \mathcal{P}) + 1}
\label{eq:ipf_av}
\end{equation}
where $|\mathcal{P}|$ is the catalog size and $N((a, v), \mathcal{P})$ counts products possessing pair $(a, v)$. Because $N((a,v),\mathcal P)\leq|\mathcal P|$, IPF is non-negative.

\paragraph{\textbf{Attribute-Value Frequency (AVF):}} AVF provides \textit{query-local discriminability} within candidate set $C_Q$:
\begin{equation}
\text{AVF}((a,v), Q) = \log\frac{|C_Q| + 1}{N((a, v), C_Q) + 1}
\label{eq:avf}
\end{equation}
A pair shared by most candidates receives a low AVF, whereas a pair unique to a few candidates receives a high AVF. Both terms are non-negative by construction.

The combined AVF-IPF weight rewards pairs that are simultaneously globally rare and locally discriminative:
\begin{equation}
W_{\text{AVF-IPF}}((a, v), Q) = \text{IPF}(a, v) \times \text{AVF}((a, v), Q)
\label{eq:avf_ipf_weight}
\end{equation}

The factor relevance score for candidate $c$ aggregates over all matched pairs:
\begin{equation}
S_{\text{attr}}(c) = \sum_{(a, v) \in \text{AVPairs}(c)\, \cap\, \text{AVPairs}(Q)} W_{\text{AVF-IPF}}((a, v), Q)
\label{eq:s_attr_av}
\end{equation}
where $\text{AVPairs}(\cdot)$ denotes the factor-value pair set of an item or query.

\paragraph{\textbf{Offline Ranking Policy.}}
\label{sec:offline_policy}
The AVF-IPF weight is a purely statistical prior: it does not know \emph{which} decision-relevant factors characterize the current query intent, nor whether visual or factor agreement should dominate. We bridge this gap with a lightweight policy that injects query-level knowledge via modulation.

The policy is a structured triple $\pi(Q) = (\rho(Q),\, o(Q),\, \mathcal{G}(Q))$: $\rho(Q) = \{\rho_{Q,a} \geq 0\}_a$ modulates the importance of each factor; $o(Q) \in \{(\textsc{vis},\textsc{attr}),(\textsc{attr},\textsc{vis})\}$ specifies the priority of the two quantized relevance axes; and $\mathcal{G}(Q) = (g_{\text{vis}},g_{\text{attr}})$ specifies their quantization granularities. The personalized score is kept as the final continuous tie-breaker rather than being included in $o(Q)$. The \textbf{trivial policy}---$\rho_{Q,a}\!\equiv\!1$, $o\!=\!(\textsc{vis},\textsc{attr})$, and the default $\mathcal{G}$---recovers the original static configuration exactly.

The policy modulates the AVF-IPF weight multiplicatively:
\begin{equation}
\widetilde{W}((a, v), Q) = \rho_{Q,a}\cdot W_{\text{AVF-IPF}}((a, v), Q)
\label{eq:mod_weight}
\end{equation}
yielding a policy-modulated factor relevance score:
\begin{equation}
\widetilde{S}_{\text{attr}}(c) = \!\!\!\sum_{(a, v) \in (\text{AVPairs}(c)\, \cap\, \text{AVPairs}(Q))}\!\!\! \widetilde{W}((a, v), Q)
\label{eq:s_attr_mod}
\end{equation}
Setting $\rho_{Q,a}\!\equiv\!1$ recovers the unmodulated score; the policy only rescales relative emphasis among decision-relevant factors for the current intent.
\paragraph{\textbf{Policy Generation.}}
\textit{Intent clustering.} Queries are grouped by category path and their predicted decision-relevant factor set. Each cluster $\mathcal{I}$ therefore represents a shared category--factor-set signature. This procedure produces approximately 12,000 clusters covering more than 95\% of daily traffic. Queries with unseen signatures fall back to the trivial production policy.

\textit{Teacher proposal and offline selection.} As in offline decision-relevant attribute learning, the policy teacher $\mathcal{T}_\pi$ uses Qwen3.7-Max. Its inputs comprise the cluster's category path, decision-relevant factors with importance scores, and AVF/IPF statistics from historical candidate sets. For each cluster, it proposes $M=8$ policy triples through schema-constrained JSON output. We evaluate the candidates on a temporally disjoint held-out replay log from day $t+1$ and retain the one maximizing exact-match nDCG@4:
\begin{equation}
\pi^*(\mathcal{I}) = \arg\max_{\pi \in \Pi_M}\; \mathrm{nDCG@4}\!\left(\operatorname{rank}(\pi,\mathcal{D}_{\mathcal{I}}^{\mathrm{eval}})\right).
\label{eq:policy_select}
\end{equation}
Candidate policies must satisfy $\rho\in[0,5]$, $o\in\{(\textsc{vis},\textsc{attr}),(\textsc{attr},\textsc{vis})\}$, and $\mathcal{G}\in\{2,3,4,5,8\}^2$. The human-annotated set used for final offline reporting is not used for policy selection.

\textit{Online serving.} At serving time, policies are stored by intent cluster in a hash map, with no online teacher invocation. A new query is assigned by its predicted category and decision-relevant factor set. The cached policy is then retrieved in expected constant time and applied through deterministic sorting, adding only one hash-map lookup to the serving path.

\paragraph{\textbf{Hierarchical Relevance Fusion.}}
\label{sec:hierarchical_fusion}
Directly injecting factor relevance into online ranking scores risks information conflicts: a high factor match can promote visually dissimilar items, degrading user trust. We therefore stratify candidates along two quantized relevance axes whose priority and granularity are controlled by the offline policy, followed by a fixed personalized tie-breaker:

\textbf{1.~Visual Relevance Grading:} Candidates are stratified into $g_{\text{vis}}$ tiers according to visual relevance $S_{\text{vis}}(c)$. Under the trivial policy, this is the primary axis.

\textbf{2.~Factor Relevance Grading:} Candidates are independently assigned to $g_{\text{attr}}$ tiers using the policy-modulated factor score $\widetilde{S}_{\text{attr}}(c)$ (Eq.~\ref{eq:s_attr_mod}), promoting exact factor matches.

\textbf{3.~Personalized Reranking:} Within the finest joint relevance tier, candidates are sorted by $S_{\text{pers}}(c)$, a function of predicted CTR and CVR; higher scores indicate greater relevance.

Let $Q^{g}(\cdot)$ map a continuous score to $g$ ordered tiers (lower means more relevant). Given the policy-selected order $o(Q)=(o_1,o_2)$, the final list is obtained by sorting candidates according to the lexicographic key
\begin{equation}
\text{key}(c) = \left( L_{o_1}(c),\; L_{o_2}(c),\; -S_{\text{pers}}(c) \right)
\label{eq:hierarchical_sort}
\end{equation}
where $L_{\textsc{vis}}(c)=Q^{g_{\text{vis}}}(S_{\text{vis}}(c))$ and $L_{\textsc{attr}}(c)=Q^{g_{\text{attr}}}(\widetilde{S}_{\text{attr}}(c))$. The negative sign on $S_{\text{pers}}$ ensures that a higher personalized score is preferred under ascending lexicographic sorting. With the trivial policy, this reduces to the fixed visual--factor--personalized hierarchy. For factor-driven intents, such as brand or specification-sensitive categories, the policy may place $\textsc{attr}$ before $\textsc{vis}$; for appearance-driven intents, visual fidelity remains first. Under either axis order, each lower-priority term can reorder only candidates tied within the tiers induced by the preceding terms.

\section{Experiments}
\begin{figure*}[t]
  \centering
  \includegraphics[width=1\linewidth]{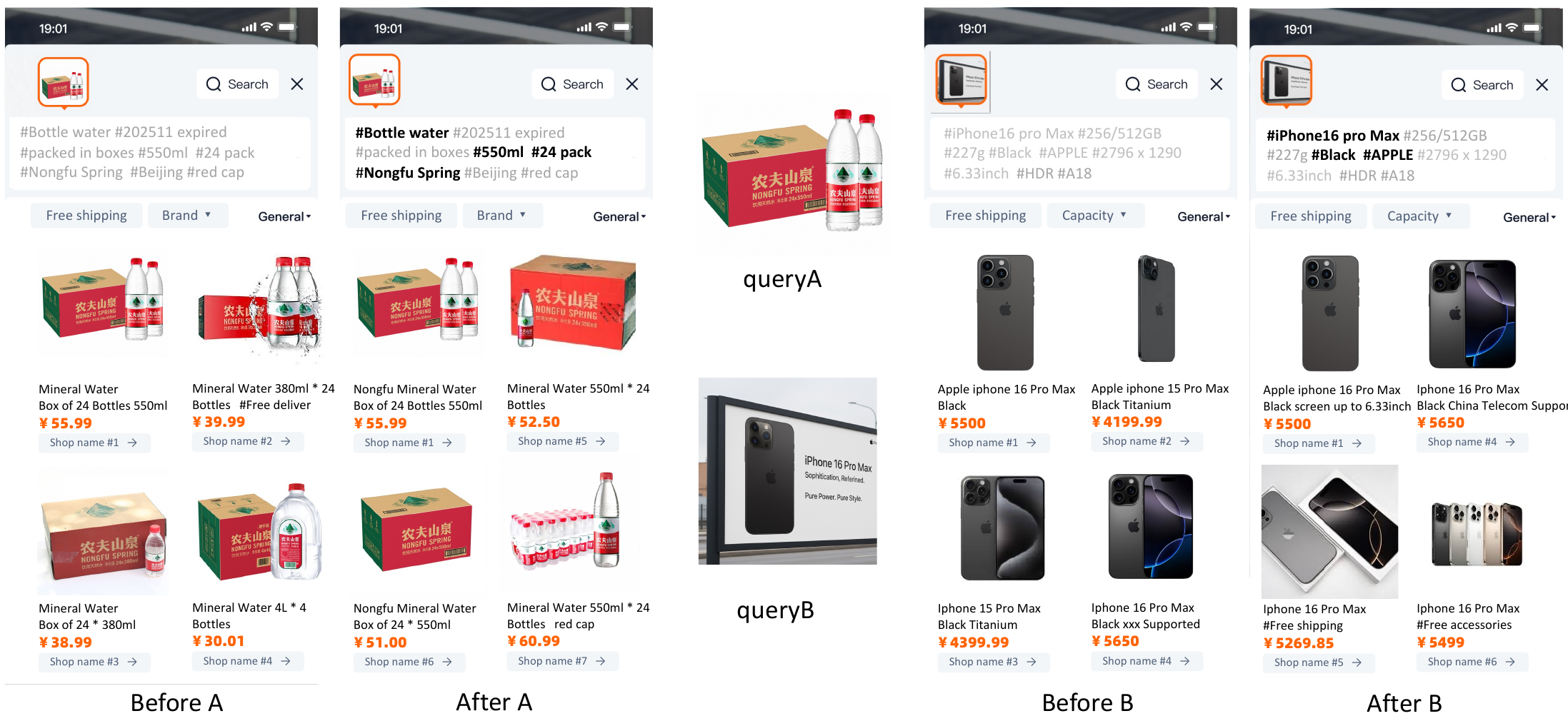}
  \caption{Qualitative comparison of the production visual-search workflow. The baseline enumerates generic attributes, whereas REM identifies behavior-informed decision-relevant attributes for factor-compatible reranking.}
\label{case_2}
\end{figure*}

We organize the evaluation according to the 3 modules of REM. For offline behavior-informed decision-relevant attribute learning, We assess merchant-curated content, behavior-grounded teacher supervision, and student generalization to behavior-sparse products. We then evaluate retrieval-augmented MLLM prediction while retaining its serving-time measurements. Finally, because heterogeneous reranking is inseparable from the production retrieval stack, we evaluate decision-relevant attribute granularity and reranking strategies through online A/B tests.
\subsection{Implementation Details and Setup}

\paragraph{\textbf{Data Preparation.}}
\textbf{(1) Multi-source product inventory.} For each product, we form a closed candidate inventory from merchant specifications, image-level predictions, and SKU-level text parsing; product and SKU titles provide item-identity and variant evidence. From one year of data spanning over 8,000 categories, we deduplicate by item ID and construct 1.1 million SFT instances (one per sampled item) containing product content, schema-valid keys, and catalog-supported values. \\
\textbf{(2) Behavior-grounded teacher evidence.} We map search and filtering interactions from the month preceding a fixed cutoff to products and aggregate key frequencies, observation counts, and exposure counts at the item and category levels. After smoothing item estimates toward category statistics, we retain at most the top 30 keys and statistics at each level. The teacher combines this demand-side evidence with the inventory to generate key-utility and value-validity targets; these behavioral statistics are privileged supervision and never student inputs. \\
\textbf{(3) Human-annotated evaluation set.} Following the production category distribution and retaining at least one item per covered category, we sample 28,900 products from over 8,000 categories. The evaluation and SFT sets are disjoint, with all SKUs of an SPU confined to one split. Annotators see titles, category paths, complete candidate inventories, and item- and category-level search and filtering evidence, but no model outputs. Ten trained professionals participate, and each sample is independently labeled by three of them to select decision-relevant keys, validate candidate values, and produce key-value sets; pairs receiving at least two votes are retained. Fewer than 15\% of samples lack a stable majority set and are reassigned, after which a trained lead adjudicates remaining disputes. These labels are collected independently of the teacher’s SFT targets and used solely for evaluation, reflecting behavior-informed human judgments rather than content-only factual annotations.
\paragraph{\textbf{Online Metrics.}}
We measure \textbf{search relevance} for Top-1 and Top-4 results across ranking buckets using four human-rated levels (Exact, Significant, Similar, and Irrelevant), and report \textbf{business metrics}: click-through rate (CTR), conversion rate (CVR), transaction count, and gross merchandise volume (GMV).

\paragraph{\textbf{Training Process.}}
All teachers use \textbf{Qwen3.7-Max} via an API with greedy decoding to generate one schema-validated target per product; invalid outputs are discarded. We fine-tune the content-only \textbf{Qwen3-30B-A3B} student~\cite{yang2025qwen3} for five epochs using AdamW, a $1\times10^{-5}$ learning rate, global batch size 64, and maximum sequence length 2,048, with greedy inference. Inputs are limited to product and SKU titles, category path, and candidate inventory. For query-side schema-constrained prediction, we train the MLLM on 4.2 million product--attribute pairs for 30 epochs on 32 NVIDIA H20 GPUs (total batch 2,048; 64 per GPU), using AdamW, a $1\times10^{-5}$ learning rate, 5\% linear warmup followed by cosine decay, and 20\% candidate masking.

\paragraph{\textbf{Deployment.}} For the ranking module, the offline stage uses 32 GPUs for \textit{t+1} production, while the online stage leverages 50 H20 GPUs to achieve 68 ms(P99) latency with a PD-disaggregated framework~\cite{pdsep}.
\begin{table}[t]
    \centering
    \fontsize{8pt}{10pt}\selectfont
    \setlength{\tabcolsep}{3pt}
    \caption{Effect of teacher-side behavioral evidence on content-only students. All variants use the same training products, teacher and student backbones, student inputs, and optimization settings; only the behavioral evidence available during teacher target generation differs. Item Beh. and Category Beh. indicate teacher access only.}
    \label{tab:offline_evidence_ablation}
    \renewcommand{\arraystretch}{1.15}
    \begin{tabular}{@{}p{0.40\linewidth}cccc@{}}
        \toprule
        Student Supervision & \shortstack{Item\\Beh.} & \shortstack{Category\\Beh.} & Key F1 & Pair F1 \\
        \midrule
        \shortstack[l]{Content-Only Teacher\\Targets} & No & No & 88.4 & 82.7 \\
        \shortstack[l]{Category-Level Teacher\\Targets} & No & Yes & 91.1 & 83.0 \\
        \shortstack[l]{Full Item--Category\\Teacher Targets} & Yes & Yes & 93.6 & 84.2\\
        \bottomrule
    \end{tabular}
\end{table}
\subsection{Offline Evaluation}
\paragraph{\textbf{Behavior-Informed Decision-Relevant Attribute Learning.}}
We compare SFT targets from teachers with progressively richer inputs: (1) a \textit{content-only teacher} given the candidate inventory, product and SKU titles, and category path; (2) a \textit{category-level teacher} additionally given category behavior; and (3) a \textit{full item--category teacher} also given item behavior. The students share training products, architecture, content-only inputs, and optimization settings, differing only in their teacher-generated targets. Behavioral statistics are available only during teacher target generation and are never included in the student inputs during training or inference.

For product $i$, $G_i^K,P_i^K$ are the reference and predicted key sets, and $G_i^P,P_i^P$ their key-value-pair sets. We compute $F_1(P,G)=2|P\cap G|/(|P|+|G|)$ on keys (Key F1) and pairs (Pair F1), reporting macro-averaged per-product scores.

Behavioral evidence informs teacher-side key-utility estimation, whereas candidate-value validity is determined from product titles, SKU structure, and cross-source consistency. This separation explains why behavioral supervision produces a larger improvement in Key F1 than in Pair F1. Because annotators also observe aggregated behavioral evidence, this benchmark evaluates agreement with behavior-informed human judgments rather than behavior-blinded factual correctness.

\paragraph{\textbf{Robustness to Behavioral Sparsity.}}
We analyze only the full student, the strongest variant in Table~\ref{tab:offline_evidence_ablation}. Within the one-month window, \textit{head} products have over 100 search exposures or over 15 filtering exposures; \textit{tail} products have positive evidence but meet neither threshold; and \textit{zero-behavior} products have no item observations but retain category evidence. With identical content-only inputs and no inference-time behavior, Pair F1 falls only 2.8 points from head (85.3) to zero-behavior products (82.5), close to the overall 84.2. Thus, content-only inference transfers to products without interaction histories, although the residual gap indicates incomplete robustness to sparsity.

\begin{table}[tb]
    \centering
    \caption{Full-student performance across product behavior strata.}
    \label{tab:sparse_generalization}
    \renewcommand{\arraystretch}{1.15}
    \begin{tabular}{lccc}
        \toprule
        Metric & Head & Tail & Zero-Behavior \\
        \midrule
        Product Share & 33\% & 49\% & 18\% \\
        Key F1 & 95.8 & 92.4 & 91.1  \\
        Pair F1 & 85.3 & 83.1 & 82.5 \\
        \bottomrule
    \end{tabular}
\end{table}

\paragraph{\textbf{Human-Annotated Search Relevance.}} On 2,500 randomly sampled online queries, human comparison of control and experimental results shows that behavior-informed decision-relevant attributes provide useful factor-level signals for exact matches (Table~\ref{tab:searchrelevance}).
\begin{table}[tb]
    \centering
    \renewcommand{\arraystretch}{1.2} 
    \caption{Human relevance evaluation on 2,500 production queries.}
    \label{tab:searchrelevance}
    \begin{adjustbox}{scale=0.95}
    \begin{tabular}{lcc}
        \toprule
        Metric & Top-1 relevance & Top-4 relevance \\
        \midrule
        Exact Match & +5.54pp & +2.79pp \\
        Overall Relevance & +2.20pp & +1.13pp \\
        \bottomrule
    \end{tabular}
    \end{adjustbox}
\end{table}
\paragraph{\textbf{Retrieval-Augmented MLLM Prediction.}}
Using the same 2,500 queries, we manually annotate candidate-set key-value references under a consistent evaluation protocol and report micro precision, recall, and F1. With the same backbone, SFT data, and optimization settings, we compare image-only input with the addition of retrieved schemas and candidates. Retrieval augmentation improves precision, recall, and F1 by 12.13, 2.46, and 7.14 points, respectively, at 33 ms additional P50 latency (Table~\ref{tab:vlm}).
\begin{table}[tb]
    \centering
    \renewcommand{\arraystretch}{1.2} 
    \caption{Micro-averaged key-value-pair prediction performance on 2,500 queries.}
    \label{tab:vlm}
    \begin{adjustbox}{scale=0.95}
    \begin{tabular}{lcccc}
        \toprule
        Input & Precision & Recall & F1 & Latency\\
        \midrule
        Image Only & 81.08\% & 83.16\% & 82.11\% & - \\
        +RAG & 93.21\% & 85.62\% & 89.25\% & +33 ms\\
        \bottomrule
    \end{tabular}
    \end{adjustbox}
\end{table}
\paragraph{\textbf{Offline Ranking Policy Ablation.}} Using human exact-match labels, we report Hit@1 (queries with an exact match ranked first), Precision@4 (exact-match fraction in the top four), and binary-relevance nDCG@4; gains are relative to the trivial policy. Axis-order adaptation ($o$) is the primary contributor (+4.0\% nDCG@4), indicating that some intents favor factor relevance over visual similarity. Adding granularity adaptation ($\mathcal{G}$) raises the gain to 6.1\%; the full policy ($\rho+o+\mathcal{G}$) reaches 6.4\%, recovers 37.8\% of the oracle gap, and exceeds the best static granularity by 1.9 points. Hence, a static hierarchy cannot replace intent-adaptive selection.

\begin{table}[tb]
  \centering
  \caption{Offline policy ablation. The trivial policy is static hierarchical relevance fusion; $\Delta$ denotes relative gain over its absolute Hit@1/Precision@4/nDCG@4 of 0.341/0.479/0.440. Oracle selects the best candidate policy per query using the same labels and upper-bounds this policy space.}
  \begin{adjustbox}{width=\linewidth}
  \renewcommand{\arraystretch}{1.1} 
  \label{tab:policy_ablation}
  \begin{tabular}{lccc}
    \toprule
    \textbf{Policy Configuration} & \textbf{$\Delta$Hit@1} & \textbf{$\Delta$Prec.@4} & \textbf{$\Delta$nDCG@4} \\
    \midrule
    Trivial (static baseline) & --- & --- & --- \\
    + $\rho$ modulation & +0.0\% & +0.0\% & +0.0\% \\
    + $o$ adaptation & +13.2\% & +2.7\% & +4.0\% \\
    + $o$ + $\mathcal{G}$ adaptation & +19.9\% & +4.0\% & +6.1\% \\
    \textbf{Full policy} ($\rho$ + $o$ + $\mathcal{G}$) & \textbf{+19.9\%} & \textbf{+4.0\%} & \textbf{+6.4\%} \\
    \midrule
    Best static $\mathcal{G}$ (grid search) & +13.2\% & +2.1\% & +4.5\% \\
    Oracle (per-query best) & +46.6\% & +16.3\% & +16.9\% \\
    \bottomrule
  \end{tabular}
  \end{adjustbox}
\end{table}

\begin{table}[!t]
  \centering
  \caption{Online A/B-test results relative to the corresponding production baseline.}
  \label{tab:abtest}
  \fontsize{7.5pt}{9.5pt}\selectfont
  \renewcommand{\arraystretch}{1.15}
  \setlength{\tabcolsep}{2pt}
  \begin{tabular*}{\linewidth}{@{\extracolsep{\fill}}p{0.39\linewidth}cccc@{}}
    \toprule
      \textbf{Module (bucket traffic)}& \textbf{CTR}&\textbf{CVR}&\textbf{Transactions}&\textbf{GMV}\\
    \midrule
   \multicolumn{5}{@{}l}{\textbf{\textit{Decision-Relevant Attrs Granularity}}} \\
    + Category-specific (5\%) & +0.23\% & +0.46\%&+0.57\%&+0.32\%  \\
    + Query-specific (5\%)& +0.71\% & +1.15\%&+1.27\%&+0.55\%  \\
    
    \midrule
    \multicolumn{5}{@{}l}{\textbf{\textit{Reranking Strategies}}} \\
   \shortstack[l]{Weighted-score\\summation (5\%)}
   & baseline & -- & -- & -- \\
    + Grading (5\%)  
   & +0.90\% & +1.21\%&+1.55\%&+3.22\%  \\
    \midrule
   \shortstack[l]{\textbf{\textit{Hierarchical Relevance}}\\\textbf{\textit{Fusion (5\%)}}} 
   & \textbf{+1.19\%} & \textbf{+1.87\%}&\textbf{+2.16\%}&\textbf{+4.11\% } \\
  \bottomrule
\end{tabular*}
\end{table}

\subsection{Online A/B Testing}

We conduct ablation and fusion A/B tests to evaluate both individual components and their integrated effectiveness in Taobao’s visual search system. Each experiment runs for two weeks and assigns 5\% of eligible traffic to its treatment bucket, which is compared with the corresponding contemporaneous production baseline. This setup provides a temporally aligned assessment of incremental online gains under normal production conditions.


\paragraph{\textbf{Decision-Relevant Attribute Granularity.}} To evaluate the effectiveness of our attribute-selection module, we conducted a three-tier online A/B test: (1) using all available attributes as the baseline, (2) pre-selecting attributes at the category level as a structured intermediate baseline, and (3) dynamically selecting a decision-relevant attribute set at the query level. As shown in Table~\ref{tab:abtest}, the category-level approach yields gains over the full-attribute baseline, with CTR +0.23\%, CVR +0.46\%, Transactions +0.57\%, and GMV +0.32\%. The query-level method further improves upon the category-level baseline across all key business metrics. These observed gains indicate that query-level selection provides more effective factor-level compatibility signals than static category-level pre-selection.

\paragraph{\textbf{Reranking Strategies.}} We compare the reranking strategies based on the query-level reasoning method. All configurations significantly boosted major business metrics (Table~\ref{tab:abtest}, $p < 0.01$).
Transactions increased by 2.16\% while PV per user decreased by 0.36\%, suggesting that the relevance improvement is associated with greater purchasing efficiency and a shorter decision process.

\paragraph{\textbf{Scalability.}} To validate scalability, we conducted a staged rollout across 5,600+ categories. Phase I (Day 1-7): 1,900 categories; Phase II (Day 8-14): remaining 3,700+. REM achieved consistent improvements, demonstrating robust cross-category generalization without per-category tuning.


  

   




\begin{figure}[!t]
\begin{promptbox}[Prompt Template for Offline Decision-Relevant Attribute Reasoning]
\fontsize{8.5pt}{10pt}\selectfont

\textbf{System:} You are an expert in e-commerce product understanding and exact-match search.\\

\textbf{Product Evidence:}\\
- Product Title: <Product\_Title>\\
- SKU Title(s): <SKU\_Titles>\\
- Category Path: <Category\_Path>\\
- Multi-source Attribute Inventory: \\
<Candidate\_Key\_Value\_Pairs>\\
- Search/Filter Evidence: <Item/Category Top-30 Keys with Statistics>\\

\textbf{Task:}\\
For each candidate key, assess its decision utility using product-content support, product-identity or SKU-variant relevance, and search and filtering evidence. For each candidate value, verify its factual validity using the product title, SKU structure, and cross-source consistency. Normalize synonymous keys, interpret implicit expressions, preserve legitimate multi-SKU values, and remove values that conflict with all product evidence.\\

\textbf{Output Format:}\\
\{\\
\hspace*{1em}"attribute\_assessments": [\\
\hspace*{2em}\{"key": "...",\\
\hspace*{3em}"content\_support": "...",\\
\hspace*{3em}"behavior\_support": "...",\\
\hspace*{3em}"values": [\{"value": "...",\\
\hspace*{4em}"valid": true/false\}],\\
\hspace*{3em}"selected": true/false\}],\\
\hspace*{1em}"decision\_relevant\_key\_values": [\\
\hspace*{2em}\{"key": "attr1",\\
\hspace*{3em}"values": ["value1"]\}],\\
\hspace*{1em}"schema\_gaps": ["..."]\\
\}
\end{promptbox}
\caption{Structured teacher prompt for offline decision-relevant attribute reasoning.}
\label{fig:prompt_template}
\end{figure}

\section{Conclusion}
We studied the gap between visual similarity and exact product matching in large-scale e-commerce search. REM addresses this gap by using behavior-informed decision-relevant attributes as a schema-grounded interface across product understanding, query interpretation, and reranking. Its dual-view teacher combines product content with item and category behavioral evidence as privileged supervision and distills the resulting decisions into a content-only student. A schema-constrained MLLM maps query images into the same attribute space, while policy-guided hierarchical fusion operationalizes their factor-level agreement for exact-match ranking.

On the 28,900 product behavior-informed human-annotated benchmark, full item--category teacher supervision improves Key F1 from 88.4 to 93.6 and Pair F1 from 82.7 to 84.2 over content-only supervision. Without accessing behavioral features at inference time, the student's Pair F1 decreases by 2.8 points, from 85.3 on head products to 82.5 on zero-behavior products. On the query side, schema-constrained retrieval augmentation improves attribute-prediction F1 from 82.11 to 89.25 with 33 ms additional latency. The full offline ranking policy improves nDCG@4 by 6.4\% relative to the trivial policy and exceeds the best static granularity by 1.9 percentage points in relative lift. End-to-end human evaluation shows a 5.54 percentage points improvement in Top-1 exact-match relevance, while online A/B testing of hierarchical relevance fusion yields gains of 1.19\% in CTR, 1.87\% in CVR, 2.16\% in transactions, and 4.11\% in GMV over weighted-score reranking.

These results suggest that interaction logs can serve as privileged offline supervision rather than mandatory serving-time features, and that a shared, executable attribute representation can connect product-side knowledge, query-side visual reasoning, and ranking decisions. Our current study relies on a closed catalog schema and platform-specific interaction logs; evaluating cross-platform transfer and robustness to evolving or incomplete schemas remains important future work.



\section*{Ethical Considerations}
This work uses platform-standard de-identified search and filtering logs only after aggregation at the item and category levels. The evidence supplied to the teacher, student-supervision pipeline, and human annotators contains aggregate product-level statistics rather than individual user histories. Human annotators are shown product content, candidate attributes, and aggregated behavioral evidence, but are not exposed to model outputs. The professional annotators receive standardized training and compensation, and the annotation workflow undergoes internal compliance review. Online experiments modify only result ordering and are evaluated through aggregate relevance and business metrics.

Interaction-derived supervision may nevertheless encode popularity and exposure biases, potentially favoring frequently interacted products or attributes. We mitigate direct serving-time dependence by distilling behavioral evidence into a content-only student and report results separately for head, tail, and zero-behavior products. Because the evaluation annotators also observe aggregated behavioral evidence, the benchmark measures agreement with behavior-informed human judgments; a complementary behavior-blinded evaluation would help disentangle shared-evidence agreement from content-only factual validity. Schema constraints and catalog-supported values reduce unsupported attribute generation, but do not eliminate errors or unequal exposure. 


\clearpage
\section*{GenAI Usage Disclosure}
\normalsize
Generative AI tools were used for language editing and manuscript-organization assistance. All research methods, implementations, experimental data, numerical results, and scientific conclusions were produced or verified by the human authors, who take full responsibility for the content of this work.
\par
{\footnotesize
\color{black}
\hypersetup{hidelinks}
\flushbottom
\setlength{\bibsep}{2pt plus 1pt minus 1pt}
\balance
\bibliographystyle{unsrtnat}
\bibliography{refs}
\par
}

\end{document}